\documentclass[preprint,showpacs,preprintnumbers,amsmath,amssymb]{revtex4}

\newcommand{\ra}{\rangle}
\newcommand{\la}{\langle}

\begin{document}

\title{Geometrical complexity of  conformations of ring polymers 
 under  topological constraints} 

\author{Miyuki K. Shimamura}
\address{
Department of Applied Physics, Graduate School of Engineering,
University of Tokyo, \\
7-3-1 Hongo, Bunkyo-ku, Tokyo 113-8656, Japan 
}

\author{Tetsuo Deguchi}
\address{
Department of Physics, Faculty of Science, Ochanomizu University,  \\
2-1-1 Ohtsuka, Bunkyo-ku, Tokyo 112-8610, Japan }


\begin{abstract} 
One measure of geometrical complexity 
of a spatial curve is 
the number of crossings in a planar projection of the curve. 
 For $N$-noded ring polymers with a fixed knot type, 
 we evaluate numerically 
 the average of the crossing number over some directions. 
We find that 
the average crossing number under the topological constraint  
 are less than that of no topological constraint 
 for large $N$. The  decrease of the geometrical complexity 
is significant when the thickness of polymers is small. 
 The simulation with or without a topological constraint also 
 shows that the average crossing number and the average 
size of ring polymers are independent measures of  
conformational complexity.    
\end{abstract}
\maketitle

\section{Introduction}

Complexity of conformations of polymer chains 
 should play  an important role in the physics of polymers. 
Various dynamical or statistical properties of entangled polymers 
in solutions or melts should be related to  
the complexity of polymer chains as space curves. 
For instance, the reptation theory takes into account 
the entanglement complexity of polymer chains 
around a polymer by assuming the tube along it \cite{Doi-Edwards}.

However, it is not trivial to investigate 
 any aspect of the complexity of conformations of polymers directly 
through computer simulation. In fact,  
it is not clear how to express the entanglement 
complexity numerically for mutually entangled polymers.  
Furthermore, it is not known how 
the complexity of polymer conformations     
should depend on the property that polymer chains 
can not cross each other.

In this paper, we discuss the geometrical complexity 
of conformations of a ring polymer 
under some topological constraint. 
As a measure of geometrical complexity 
of a spatial curve, we consider    
the number of crossings in a planar projection 
of the curve, and take the  average over some directions 
\cite{Janse92,Orlandini}.  
Through computer simluation, we evaluate  
the average crossing number  
for $N$-noded ring poymers with fixed knots, 
and discuss their behaviors with respect to $N$.     
We thus make it clear how the topological constraints 
can modify the geometrical complexity. Here we note that 
the condition that a ring  polymer never cross with itself   
corresponds to a topological constraint on it,  
as far as its statistical properties are concerned.   
Furthermore, we plot the graphs of 
 the mean square radius of gyration versus 
 the average crossing number,  
for the SAPs with fixed knots.  
Then we find that the average crossing number and the mean-squared 
gyration radius behave independently under the topological constraints. 
It is thus suggested that the relation 
between the average sizes of ring polymers 
and the geometrical complexity of the conformations should be 
non-trivial. 

Let us explain  the average crossing number, more precisely.  
The {\it writhe} of a linear or ring  polymer is defined by the average 
of the number of signed crossings appearing 
in a projection of the curve over all directions.  
As a simplified version of the writhe, 
Janse van Rensburg $et.al$ introduced 
{\it the number of crossings} \cite{Janse92}.
It is defined by the number of unsigned crossings in a projection. 
Then, {\it the average number of crossings} is defined by 
the number of crossings averaged over all possible projections. 
There are several numerical or theoretical studies 
on the average crossing number 
\cite{Arteca,Lai,Vilgis,Grass}.

The average crossing number is  also related to ideal knots. 
For a knot, the ideal knot is given by its tightest geometric configuration 
\cite{KatritchNature,KatritchBook,GrosbergPRE}.
Katritch {\it et. al} have obtained  
the average crossing number for the ideal knots 
up to 11 essential crossing numbers. 
The average crossing number should be useful for flexible DNA knots 
in thermal equilibrium \cite{KatritchNature}. 
Furthermore, it should also be useful 
for statistical or dynamical studies on 
knotted ring polymers \cite{Lai,GrosbergBook,PRE2002R} .

\section{Methods of simulation}

We consider self-avoiding polygons (SAPs) consisting of $N$ 
rigid impenetrable cylinders 
of unit length and radius $r$. 
There is no overlap allowed between any pair of non-adjacent 
cylindrical segments, 
while next-neighboring cylinders may overlap each other. 
We call them cylindrical SAPs, for short.
A large number of cylindrical SAPs can  be constructed 
by the cylindrical ring-dimerization method \cite{PLA}.
The method is based on the algorithm of ring dimerization \cite{Chen}, 
and  useful for generating long SAPs systematically. 
In this paper, we have constructed $M$=$10^4$ samples of $N$ cylindrical segments
 with radius $r$, where $N$ is from 20 to 1000 and  $r$ is from 0.0 to 0.07.
We determine the number $M_K$ of polygons with a knot $K$, enumerating 
such polygons that have the same set of values of the two invariants: 
the determinant of the knot $\Delta_K(-1)$ and the Vassiliev invariant $v_2(K)$ 
of the second degree.

The mean value $A$ of the average crossing number for a set of SAPs 
is defined by $A= \sum_{i=1}^{M} A_i/M $. Here, $A_i$ denotes 
the average crossing number of the $i$th polygon.  
The mean value of the average crossing number 
for a set of the SAPs with a  knot $K$
is given by  $A_K= \sum_{i=1}^{M_K} A_{K,i}/M_K $,
where $A_{K,i}$ denotes the average crossing number of the $i$th SAP  
having the knot $K$. Thus, $A$ is given by the  average of $A_K$ 's over all knots. 
\par 
In the simulation, we have  obtained  
the average crossing number $A$ 
 and $A_K$'s for the trivial and trefoil knots 
 in the range of $N$ from 20 to 1000.  Here,  
we evaluate the average crossing number  
by taking the average over the  $x$, $y$ and $z$ directions.

The mean square radius of gyration $R^2$ for  SAPs with $N$ nodes  
is given by $R^2= \sum_{n,m=1}^{N} \la (\vec{R_n}-\vec{R_m})^2 \ra/2N^2$.  
Here ${\vec R}_n$ is the position vector of the $n$th node,  
and the symbol $\la \cdot \ra$ denotes 
the average over $M$  polygons generated. 
For a  knot $K$, we define the mean square radius of gyration 
$R_K^2$ for SAPs of the knot $K$ by 
 $R_K^2=\sum_{i=1}^{M_K} R_{K,i}^2/M_K$ \cite{PRE2001R,PRE2002}. 
Here, $R_{K,i}^2$ denotes the mean square radius of gyration for 
 the $i$ th SAP in the $M_K$ polygons of the knot $K$. 
In terms of $R_K^2$'s, 
$R^2$ is given by the average over all knots: 
 $R^2=\sum_{K}M_K R_K^2/M$. 

\section{Average crossing number under a topological constraint}

We now discuss 
 the $N$ dependence of the average crossing number $A_K$ 
 under the topological constraint of a  knot $K$. 
The double-logarithmic graph of $A_K$ versus $N$ is given in Fig. 1 
for the cylindrical SAPs of cylinder radius $r=0.003$, for 
the trivial and trefoil knots.  
The graphs can be approximated by 
 some straight lines in some large $N$ region.

Let us consider the ratio of $A_K$ to $A$ for the two knots. 
In Fig. 2, the ratio $A_K/A$ versus $N$ is plotted 
in a double logarithmic scale for the trivial and trefoil knots. 
The graph of the trivial knot has a concave curve: 
the ratio $A_{triv}/A$ is almost constant with respect to $N$ for small $N$ 
and then decreases with a larger gradient for large $N$.  
On  the other hand,  the graph of the trefoil knot 
suggests that $A_{tre}/A$ 
could be roughly approximated by a power of $N$ for some 
finite values of $N$.

In Fig. 2, the ratios $A_K/A$'s for the two knots 
become less than 1.0 when $N$ is large, 
 i.e., the $A_K$'s are less than 
the average crossing number $A$ averaged over all knots 
for large $N$. Thus, 
the topological constraints make the 
conformations of ring polymers simpler with respect to  
 the geometric complexity, when $N$ is large enough.  
The reduction of conformational complexity may be related to 
entropic repulsion arising from the topological constraints.

For the trivial knot, 
 $A_{triv}/A$ is smaller than 1.0 in the whole range of $N$,  
while for the trefoil knot 
$A_{tre}/A$ is larger than 1.0 for  $N < 200$, 
and smaller than 1.0 for $N > $ 300.  
Thus, with respect to the average crossing number,  
the $N$-noded ring polymers with the trivial knot 
are  less complex for any $N$ 
than those of no topological constraint ,   
while those of the trefoil knot are more complex 
for small $N$  and less complex for large $N$ 
than those of no topological constraint.

The average crossing number of the trivial knot, $A_{triv}$,   
is smaller than that of the trefoil knot, $A_{tre}$, 
through the whole range of $N$. 
It suggests that the more complicated knot should have 
the larger value of the average crossing number, at least for finite $N$.  
The tendency is also seen in the data for the different values of $r$.
As $N$ increases, however, the average crossing numbers 
$A_K$'s of the two knots gradually become close to each other  
 and they become almost 
 the same value when $N$ is very large.

 The average crossing number $A_K$ of a knot $K$   
 have  similar properties with the inverse of 
the mean-squared gyration radius $R^2_K$ 
of ring polymers with the knot $K$. 
In our previous work \cite{PRE2001R,PRE2002}, 
it is shown  that for some random polygons and cylindrical SAPs, 
the double-logarithmic graph of 
the ratio $R_K^2/R^2$ versus $N$ is given by a downward convex curve 
and is larger than 1.0 for the trivial knot, 
while for the trefoil knot the graph is given by a straight line 
and the ratio $R_K^2/R^2$ is smaller than 1.0 for small $N$ and 
larger than 1.0 for large $N$. 


We now consider possible asymptotic behaviors 
of the average crossing number.  
Let us review some known results on the large $N$ 
behavior of $N$-step self-avoiding walks (SAWs). 
The average crossing number can also be defined 
for SAWs, and we denote it by $A_{\rm SAW}$ . 
In Ref. \cite{Grass}, it is discussed that 
for asymptotically large $N$, 
the mean value $A_{\rm SAW}$ of the average crossing number of 
self-avoiding walks (SAWs) is given by $A_{\rm SAW}/N = a - b N^{1 - 2 \nu}$ 
 with some constants $a$ and $b$. 
Here $\nu$ is given by the exponent of the average size of SAWs.      
 We also note that in Ref. \cite{Orlandini}, the large $N$ behavior of 
  $A_{\rm SAW}$ is approximated by a power of $N$, 
$A_{\rm SAW} \sim N^{\mu_{SAW}}$, with the effective exponent: 
$\mu_{SAW}$ =1.122 $\pm$ 0.005.

In order to illustrate some $r$ dependent properties of $A_K$,  
let us introduce an asymptoic expansion for the ratio $A_K/A$ versus $N$.  
Based on the asymptotic expansion of $A_{\rm SAW}$ 
in Ref. \cite{Grass}, 
 we assume the asymptotic expansion with repsect to $N$ as follows: 
$A_K/N = a_K - b_K \,  N^{1-2 {\nu_K}}$  for any knot $K$. 
Here $a_K$ and $b_K$ are fitting parameters.  
 From the simulations \cite{PRE2002,PRE2001R}, 
 we may assume that $\nu_K=\nu$.    
 For the ratio $A_K/A$, we  thus have  the following formula:  
$A_K/A =  \alpha_K \, \left( 1- \beta_K \,  N^{1-2 \nu} \right)$.  
Here, $\alpha_K$ and $\beta_K$ are  fitting parameters 
corresponding to $a_K/a$ and $b_K/a_K - b/a$, respectively. 
We apply it to the data of $A_K/A$ 
of the $N$-noded SAPs with $N$ larger than some cut-off and  
for the different values of $r$.

Let us discuss the best estimates of $a_K/a$ plotted 
against cylinder radius $r$ in Fig. 3. 
When $r$ is small, the ratio $a_K/a$  
becomes smaller than 1.0 both for the trivial 
and trefoil knots. 
This is consistent with the observation of Fig. 2 
that the ratio  $A_K/A$  for $r=0.003$ 
decreases against $N$ 
and is smaller than 1.0 for large $N$. 
In Fig. 3,  the  ratio $a_K/a$ increases monotonically 
with respect to cylindrical radius $r$,  
and it becomes close to the value 1.0 at some  
large  value of $r$ .  If the ratio $A_K/A$ becomes 
1.0, then there is no topological effect 
on the average crossing number. 
On the other hand, if $A_K/A$ is less than 
1.0, then it may be  a consequence of the topological constraint.  
Thus, the behavior that the graph of $a_K/a$ versus 
 $r$ increases upto 1.0  suggests   
that topological constraints on ring polymers 
make their conformations simpler for small $r$,  
while  the topological effect becomes weak 
for large $r$.

Let us consider again the $r$ dependence of $A_K$ that the graph of $a_K/a$ versus 
 $r$ increases upto 1.0 for the two knots,  as shown in Fig. 3.   
A very similar  behavior has also been observed for the case of 
the ratio $R_K^2/R^2$ of the mean squared gyration radii 
of the cylindrical SAPs with radius $r$.     
In Refs. \cite{PRE2001R,PRE2002}, it has been shown 
for the cylindrical SAPs 
that a topological constraint on a ring polymer 
gives effective expansion to it, i.e., 
the ratio $R_K^2/R^2$ becomes larger than 1.0 for large $N$.   
Furthermore, the effective expansion 
is significant when the cylinder radius $r$ is small;  
the large $N$ limit of $R_K^2/R^2$ decreases to 1.0 as 
$r$ increases. 

Summarizing the simulation results, we may conclude 
that the effect of topological constraints should be significant 
when the ring polymer is thin, both for the average crossing 
number $A_K$ and the mean square radius of gyration $R_K^2$.

\section{The mean squared gyration radius 
and the geometric complexity}

Let us discuss that  the relation between the average crossing number and 
the average size of ring polymers  should be nontrivial. 
In Fig. 4,  the mean square radii of gyration $R^2$ and $R_K^2$'s 
are plotted against the average crossing numbers $A$ and $A_K$'s 
in a double logarithmic scale 
for the cylindrical SAPs of $r$=0.003. 
The estimates of $R^2$ and $R^2_K$'s in Fig. 4 
are different for any given value of the average crossing numbers.   
Thus, it follows that 
the mean-squared gyration radius and the average crossing number 
are independent quantities describing some geometric properties of 
conformations of ring polymers.

 With the same average crossing number given, 
 $R_{triv}^2$ and $R_{tre}^2$ are larger than $R^2$ 
 for the cylindrical SAPs, as shown in Fig. 4. 
Thus,  we may say that the topological constraints make  
the average sizes of ring polymers larger  
with respect to the average crossing number.   
Among the $R_K^2$'s,  $R_{triv}^2$ is larger than $R_{tre}^2$ 
in Fig. 4. It is thus suggested that 
the more complicated knot should have  
the smaller radius of gyration, as far as 
 some finite values of $A_K$'s are concerned.

When the average crossing number is very large, 
the mean-squared gyration radius $R^2$ (or $R_K^2$) can 
be approximated by some power of $A$ (or $A_K$):  
$R^2 \sim A^{\gamma}$ for the average over all knots;   
$R_K^2 \sim A_K^{\gamma_K}$ for the trivial and trefoil knots.
Applying the power law approximation to the data of $N \ge 100$, 
we obtain the following estimates of the effective exponents:   
$\gamma=0.860 \pm0.001$, $\gamma_{triv}= 0.952 \pm 0.004$ 
 and $\gamma_{tre} = 1.069 \pm 0.006$. 
The graphs of Fig 4 suggests that 
the effective exponent $\gamma_K$ 
should  be independent of the knot type. 
Similar results are obtained also for the SAPs 
with the different values of $r$. 

\begin{acknowledgements} 
We would like to thank  Prof. K. Ito  for  helpful discussions.  
\end{acknowledgements}





%
\begin{figure}
\caption{The average number of crossings with a knot type $K$ $A_K$ versus $N$ for $r$=0.003. 
Numerical estimates of $A_K$ for $K$=trivial and trefil knots are shown by
closed triangles and squares, respectively.}
\end{figure}
%
\begin{figure}
\caption{Double-logarithmic plots of the ratio $A_K/A$ versus $N$ for cylindrical SAPs
with $r$=0.003.
$A_{triv}/A$ and $A_{tre}/A$ are shown by closed triangles and squares, respectively.}
\end{figure}

%
\begin{figure}
\caption{The ratio $a_K/a$ versus cylinder radius $r$ for cylindrical SAPs. 
The values of $a_{triv}/a$ and $a_{tre}/a$ are shown by closed triangles
and squares, respectively.}
\end{figure}

%
\begin{figure}
\caption{The mean-square radius of gyration $R_K^2$ versus the average number of crossings
 $A_K$ of cylindrical SAPs for $r$=0.003.
Numerical estimates for $K$=trivial knot, $3_1$ are shown by closed triangles 
and squares, respectively.
Data of $R^2$ versus $A$ are shown by closed circles.
$N$ are given by 51, 151 and 100$j$+1 for $j$=1,$\cdots$,10.}
\end{figure}


\begin{thebibliography}{[99]}

\bibitem{Doi-Edwards} M. Doi and S.F. Edwards, 
{\it The Theory of Polymer Dynamics}  (Clarendon Press, Oxford, 1986).  

\bibitem{Janse92} E. J. Janse van Rensburg, 
D. A. Sumners, E. Wasserman and S. G. Whittington,  
J. Phys. A: Math. Gen. {\bf 25} (1992) 6557.


\bibitem{Orlandini} E.Orlandini {\it et.al}, J.phys.A: Math Gen. {\bf 27} (1994)L333.

\bibitem{Arteca} G.A. Arteca, Phys. Rev. E {\bf 49} (1994) 2417.

\bibitem{Lai} J-Y Huang and P-Y Lai, Phys. Rev E {\bf 63} (2001) 021506. 

\bibitem{Vilgis} A.L. Kholodenko and T.A. Vilgis, Phys. Rep. {\bf 298} (1998) 251.

\bibitem{Grass}P. Grassberger, J. Phys. A : Math. Gen. {\bf 34} (2001) 9959.

\bibitem{KatritchNature}
V. Katritch, J. Bednar, D. Michoud, R. G. Scharein, J. Dubochet, and A. Stasiak,
Nature (London) {\bf 384}, 142 (1996)

\bibitem{KatritchBook}
A. Stasiak, J. Dubochet, V. Katritch, and P. Pieranski, 
in {\it Ideal Knots} edited by A. Stasiak, V. Katritch and L. H.
Kauffman (World Scientific, Singapore, 1998) pp. 1-19. 

\bibitem{GrosbergPRE}
A. Yu. Grosberg, A. Feigel, and Y. Rabin, Phys. Rev. E {\bf 54}, 6618 (1996).

\bibitem{GrosbergBook}
A. Yu. Grosberg, in {\it Ideal Knots} edited by A. Stasiak, V. Katritch and L. H.
Kauffman (World Scientific, Singapore, 1998) pp. 129-142. 

\bibitem{PRE2002R} M.K. Shimamura and T. Deguchi, 
Phys. Rev. E {\bf 66}, R040801(2002).


\bibitem{PLA} M.K. Shimamura and T. Deguchi,
Phys. Lett. A  {\bf 274}, 184 (2000). 


\bibitem{Chen} Y.D. Chen, J. Chem. Phys. {\bf 74}, 2034 (1981) ; J. Chem.
Phys. {\bf 75}, 2447 (1981); J. Chem. Phys. {\bf 75}, 5160 (1981).


\bibitem{PRE2001R} M.K. Shimamura and T. Deguchi, 
Phys. Rev. E {\bf 64}, R 020801 (2001).

\bibitem{PRE2002} M.K. Shimamura and T. Deguchi, 
Phys. Rev. E {\bf 65}, 051802 (2002).



\end{thebibliography}
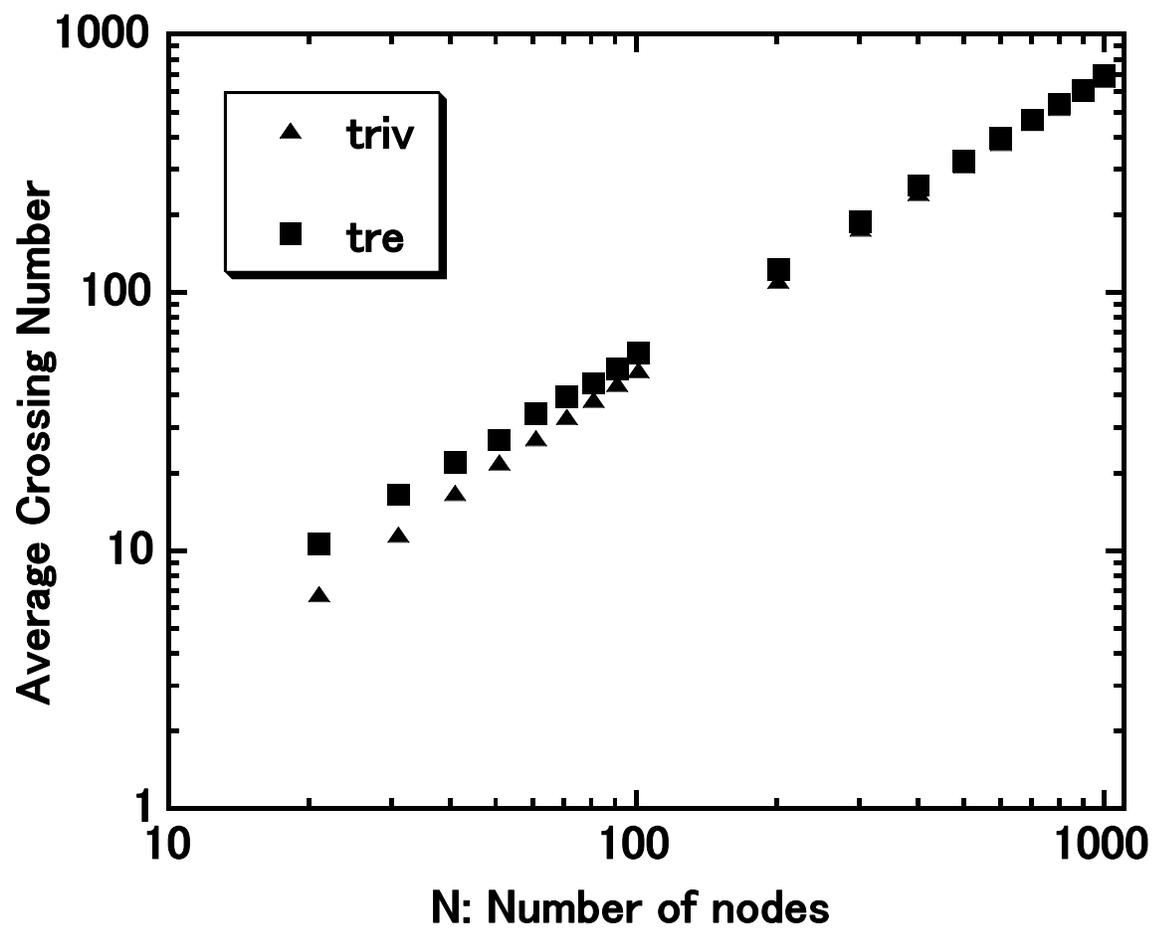
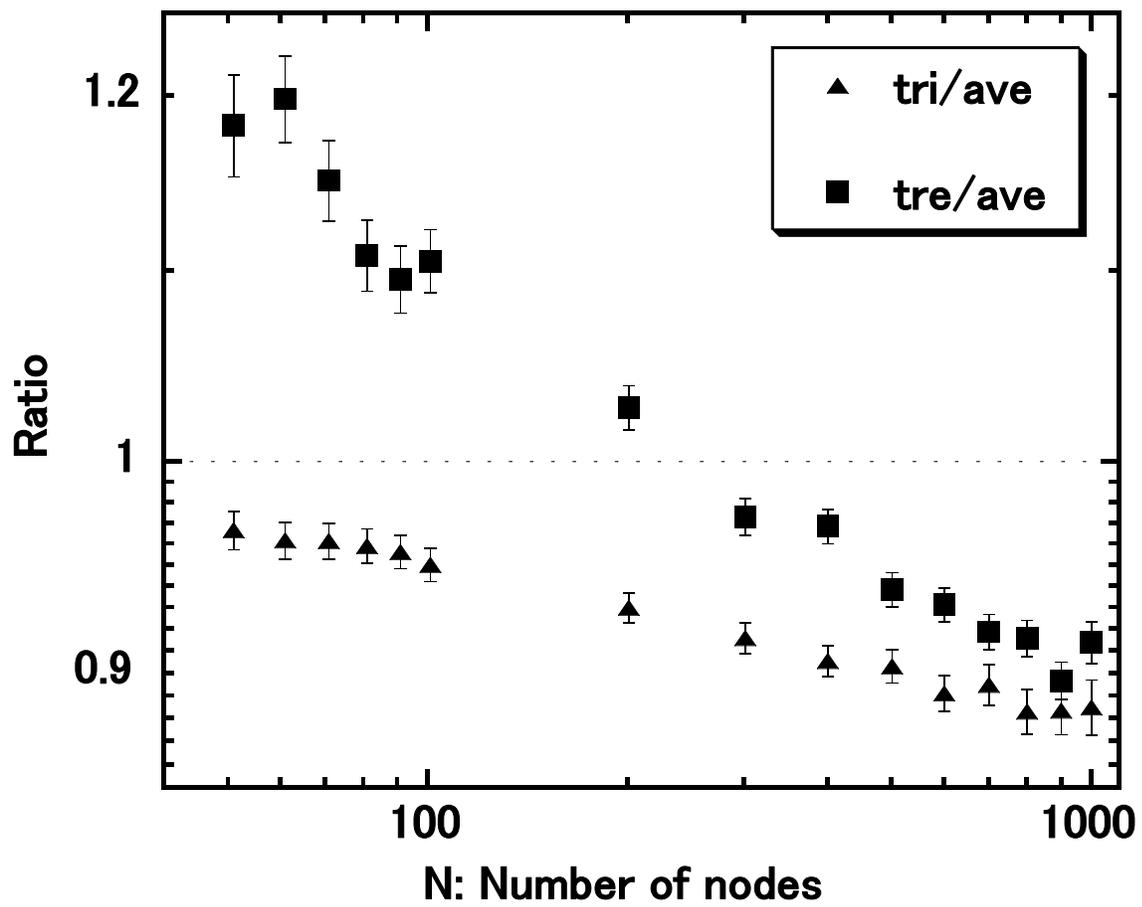
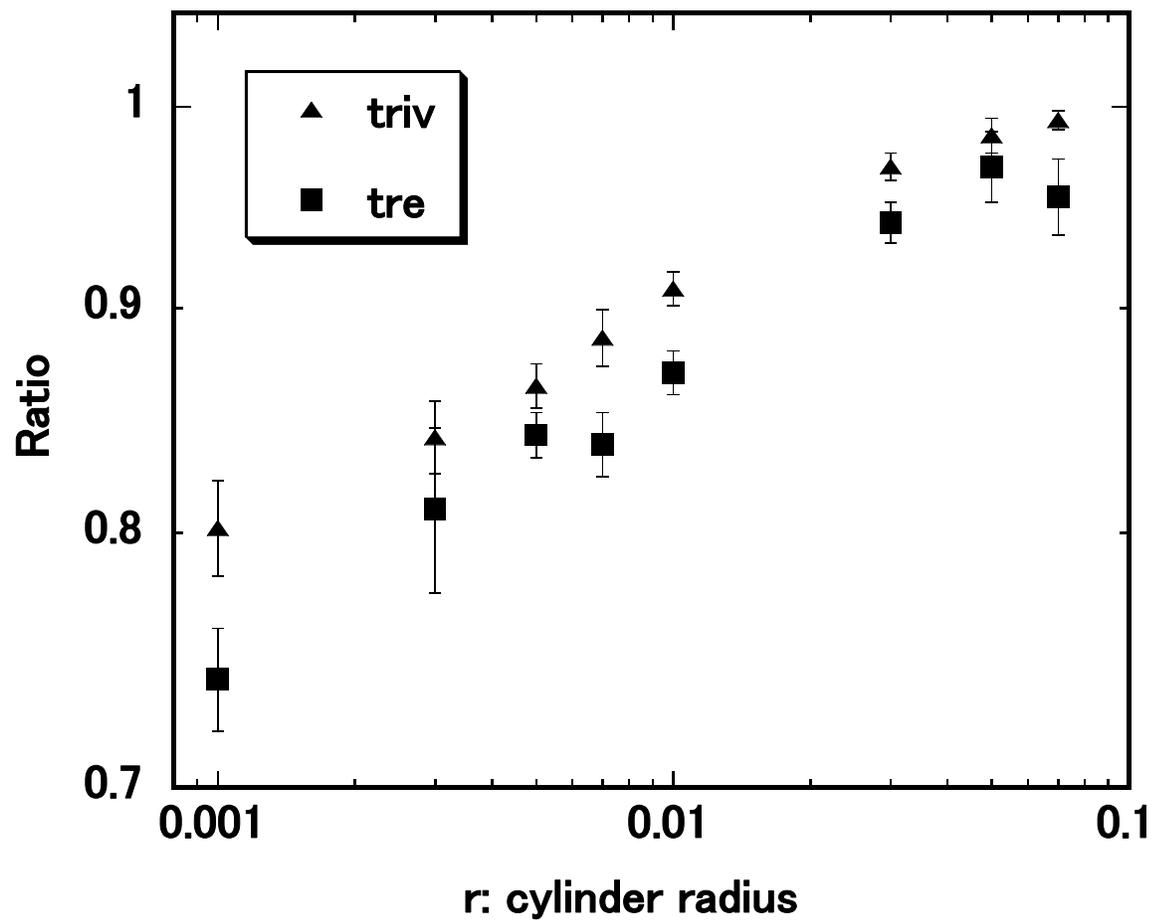
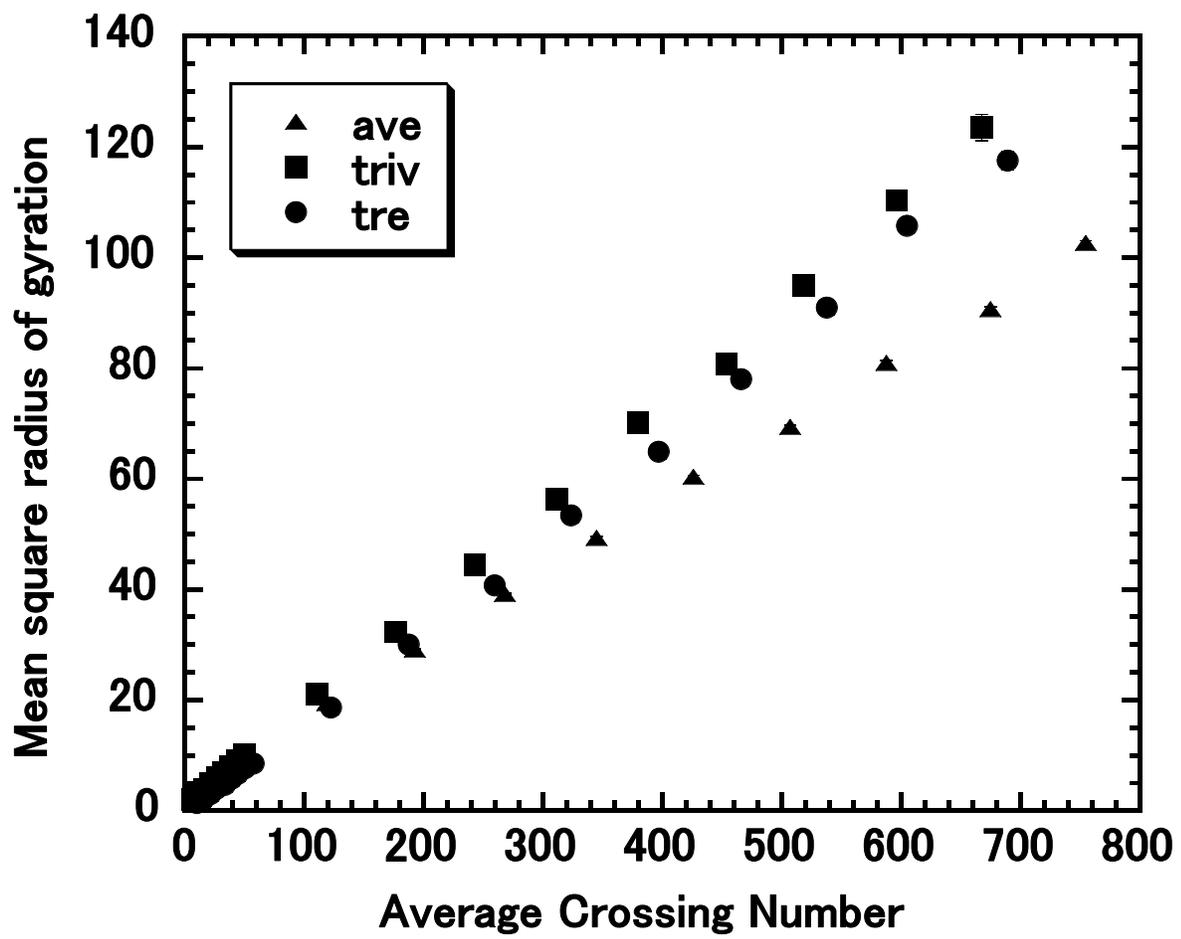
\end{document}